\documentclass[sigconf]{acmart}

\usepackage{multirow}
\usepackage{graphicx}
\usepackage{subfigure}
\AtBeginDocument{%
  \providecommand\BibTeX{{%
    \normalfont B\kern-0.5em{\scshape i\kern-0.25em b}\kern-0.8em\TeX}}}

\setcopyright{acmcopyright}
\copyrightyear{2018}
\acmYear{2018}
\acmDOI{XXXXXXX.XXXXXXX}

\acmConference[Conference acronym 'XX]{Make sure to enter the correct
  conference title from your rights confirmation emai}{June 03--05,
  2018}{Woodstock, NY}
\acmPrice{15.00}
\acmISBN{978-1-4503-XXXX-X/18/06}

\begin{document}

\title{Evaluating the Impact of Tiled User-Adaptive Real-Time Point Cloud Streaming on VR Remote Communication}


\author{Shishir Subramanyam}
\email{s.subramanyam@cwi.nl}
\affiliation{%
 \institution{CWI}
 \city{Amsterdam}
 \country{the Netherlands}
}

\author{Irene Viola}
\email{irene@cwi.nl}
\affiliation{%
 \institution{CWI}
 \city{Amsterdam}
 \country{the Netherlands}
}

\author{Jack Jansen}
\email{jack.jansen@cwi.nl}
\affiliation{%
 \institution{CWI}
 \city{Amsterdam}
 \country{the Netherlands}
}

\author{Evangelos Alexiou}
\email{evangelos.alexiou@cwi.nl}
\affiliation{%
 \institution{CWI}
 \city{Amsterdam}
 \country{the Netherlands}
}
\author{Alan Hanjalic}
\email{a.hanjalic@tudelft.nl}
\affiliation{%
 \institution{TU Delft}
 \city{Delft}
 \country{the Netherlands}
}

\author{Pablo Cesar}
\email{p.s.cesar@cwi.nl}
\affiliation{%
\institution{CWI}
 \city{Amsterdam}
 \state{the Netherlands}\\
 \institution{TU Delft}
 \city{Delft}
 \country{the Netherlands}
}

\renewcommand{\shortauthors}{}

\begin{abstract}
Remote communication has rapidly become a part of everyday life in both professional and personal contexts. However, popular video conferencing applications present limitations in terms of quality of communication, immersion and social meaning. VR remote communication applications offer a greater sense of co-presence and mutual sensing of emotions between remote users. Previous research on these applications has shown that realistic point cloud user reconstructions offer better immersion and communication as compared to synthetic user avatars. However, photorealistic point clouds require a large volume of data per frame and are challenging to transmit over bandwidth-limited networks. Recent research has demonstrated significant improvements to perceived quality by optimizing the usage of bandwidth based on the position and orientation of the user's viewport with user-adaptive streaming. In this work, we developed a real-time VR communication application with an adaptation engine that features tiled user-adaptive streaming based on user behaviour. The application also supports traditional network adaptive streaming. The contribution of this work is to evaluate the impact of tiled user-adaptive streaming on quality of communication, visual quality, system performance and task completion in a functional live VR remote communication system. We perform a subjective evaluation with 33 users to compare the different streaming conditions with a neck exercise training task. As a baseline, we use uncompressed streaming requiring ca. 300Mbps and our solution achieves similar visual quality with tiled adaptive streaming at 14Mbps. We also demonstrate statistically significant gains to the quality of interaction and improvements to system performance and CPU consumption with tiled adaptive streaming as compared to the more traditional network adaptive streaming.
\end{abstract}




\maketitle

\section{Introduction}
Remote communication and collaboration has rapidly become a necessity in a globalized and connected world. Video conferencing applications have become ubiquitous in everyday life in both professional and personal environments. Notwithstanding their popularity, it is estimated that travel for the purpose of in-person communication is responsible for roughly eight percent of US energy consumption~\cite{energyspentf2f}. With recent events like the Covid-19 global pandemic there is an increased need for applications that can deliver a greater sense of co-presence and mutual sensing of emotions in remote communication. Current video conferencing solutions have clear limitations in this regard~\cite{jie:photosharingCHI,VRMeeting:ieeespectrum,McNamara2021Nonverbal:ZoomFatigue,Yassein:SocialPresence:DesignSpace}. 

Immersive Virtual Reality (VR) applications offer an increased sense of presence and immersion. These applications have emerged as a promising alternative for remote communication and telepresence~\cite{3dteleimmersion:viewcast,Fuchs:Telepresence:General, mekuria:tele, jack, simon:MCU, roomalive:orig, google:starline,microsoft:holoportation}. They allow users to employ both verbal and non-verbal communication in a shared virtual space. In such applications, users can be embodied in the virtual space either using avatars or real-time photorealistic 3D reconstructions typically using depth sensors. Previous work in the field has demonstrated that realistic user reconstructions improve immersion and communication~\cite{mekuria:virtualroom, latoschik:realisticavatars} as compared to avatars.

Among the different 3D formats, point clouds have emerged as a popular representation for user reconstructions as they are relatively easy to acquire in real-time using consumer depth sensors~\cite{simon:MCU,jack}. This format represents the object's geometry as an unorganized collection of surface point coordinates with additional  attributes such as color provided at each point location. They are generally resilient to noise and do no incur an additional computational overhead to triangulate mesh faces making them suitable for real-time applications. Owing to their unorganized nature they are also easy to partition into non-overlapping segments. However, photorealistic point clouds present a large volume of data per frame that requires real-time compression to transmit over bandwidth-limited networks~\cite{Jeroen:MM:PCAdaptiveStreaming, Liu:PCStreamingGeneral}. In order to alleviate this requirement recent research has looked into adapting the point cloud stream to the user's viewport location and orientation in order to optimize how the available bandwidth is spent. This is done by prioritizing objects or surfaces facing the viewer and reducing the wastage of bandwidth on surfaces or objects that are occluded or outside the viewport ~\cite{christian:mmsys18:pcstreaming,phil:mmsys18:pcstreaming,Jeroen:MM:PCAdaptiveStreaming,cwi:mm20,Gul:IBC:CloudVolumetricVideo, Son:SplitRendering:SIGGRAPH}. However this approach to adaptive streaming has not yet been evaluated for live communication with real-time point cloud reconstructions.

In this work, we implement and evaluate a two user social VR system with real-time adaptive streaming of user reconstructions shown in figure~\ref{fig:arch}. We set out to assess the impact of tiled adaptive streaming on the quality of communication, visual quality and subjective task related performance. We constructed a social VR pipeline extending the system proposed in~\cite{jack} with network adaptive and tiled adaptive streaming. Two confederate users were recruited and trained to play the role of a trainer in every experiment session while 33 users (16 females, 17 males) were recruited to play the role of trainee. The participants were asked to learn and perform three neck exercises during the session. The contribution of this work is to evaluate and compare tiled adaptive streaming (\textit{TA}) with traditional network adaptive streaming (\textit{NA}) and baseline uncompressed streaming in a functional live VR remote communication system. We propose and employ a novel evaluation methodology using a training task to perform the assessment. We address the following four research questions for VR remote communication: 
\begin{itemize}
\item {\texttt{R1}}: How does tiled user-adaptive point cloud streaming impact quality of interaction/ quality of communication \textit{(QoI)}?
\item {\texttt{R2}}: How does tiled user-adaptive point cloud streaming impact the experience of performing a training task?
\item {\texttt{R3}}: How does tiled user-adaptive point cloud streaming impact the perceived quality of remote user reconstruction?
\item {\texttt{R4}}: What is the computational overhead of using tiled adaptive point cloud streaming? 
\end{itemize}

From our results, we observe statistically significant improvements to QoI (\texttt{R1}). We observe no statistically significant change to task experience (\texttt{R2}). We observe significant improvements to visual quality (\texttt{R3}) and at 14Mbps we observe similar visual quality to uncompressed point clouds streamed at ca. 300Mbps. We also see a reduction in CPU utilization and an improvement to playback performance (\texttt{R4)}. We validated the communication system and checked that the training task provides coherent results in the evaluation.

\section{Related Work}
\label{sec:relatedWork}
\subsection{VR Remote Communication using Point Clouds}
Advances in low-latency streaming and volumetric point cloud delivery mechanisms have led to the emergence of novel teleimmersion systems that allow distributed remote users to communicate as themselves in a shared 
environment with realistic user reconstructions. 
Microsoft released the RoomAlive Toolkit for creating interactive Augmented Reality (AR) experiences \cite{roomalive:magic,roomalive:orig}.
Mekuria et al. proposed a teleimmersive system that blends avatar representations and photo-realistic reconstructions of users in a shared virtual environment~\cite{mekuria:virtualroom}. Cernigliaro et al. propose a point cloud multi-point control unit for optimizing holo-conferencing systems~\cite{i2cat:mcu}. Gunkel et al. introduced VRComm~\cite{simon:MCU}, a web based social VR communication system using photo-realistic user reconstructions that was evaluated using both simulations and subjective studies. Jansen et al.~\cite{jack} proposed a pipeline for volumetric videoconferencing using low latency DASH with photo-realistic point cloud user reconstructions. \textbf{In this work}, we extend this system design with network adaptive \textit{(NA)} and tiled user-adaptive \textit{(TA)} streaming. We transmit tiled point cloud user reconstructions at fixed target bitrates to assess the experience without the influence of a volatile network.

\begin{table*}[t]
\caption{System Setup}
\label{tab:systemSetup}
\begin{tabular}{|c|l|lll|}
\hline
\multirow{5}{*}{Hardware}                & HMD                                   & \multicolumn{3}{l|}{HTC   Vive Pro Eye}                                                      \\ \cline{2-5} 
                                         & CPU                                   & \multicolumn{3}{l|}{Intel(R) Core(TM) i7-7700K   CPU @ 4.20GHz (8 CPUs), $\sim$4.2GHz}       \\ \cline{2-5} 
                                         & GPU                                   & \multicolumn{3}{l|}{NVIDIA GeForce GTX 1080 Ti}                                              \\ \cline{2-5} 
                                         & Memory                                & \multicolumn{3}{l|}{32768MB RAM}                                                             \\ \cline{2-5} 
                                         & Depth Sensors                         & \multicolumn{3}{l|}{3 x Azure Kinect DK}                                                     \\ \hline
\multirow{3}{*}{Display   Parameters}    & Resolution                            & \multicolumn{3}{l|}{1440 x 1600 pixels per eye}                                              \\ \cline{2-5} 
                                         & Application Target Framerate          & \multicolumn{3}{l|}{90 Hz}                                                                   \\ \cline{2-5} 
                                         & Point Cloud Playback Target Framerate & \multicolumn{3}{l|}{15 Hz}                                                                   \\ \hline
\multirow{5}{*}{Fixed System Parameters} & Audio Codec                           & \multicolumn{3}{l|}{Ogg Speex 48 KHz}                                                        \\ \cline{2-5} 
                                         & Point Cloud Codec Configurations      & \multicolumn{1}{l|}{Octree Depth 6}   & \multicolumn{1}{l|}{Octree Depth 7} & Octree Depth 9 \\ \cline{2-5} 
                                         & Kinect Depth Configuration            & \multicolumn{3}{l|}{NFOV unbinned 640x576}                                                   \\ \cline{2-5} 
                                         & Kinect Color Configurations           & \multicolumn{3}{l|}{1280x720}                                                                \\ \cline{2-5} 
                                         & Point Cloud Capture                   & \multicolumn{3}{l|}{ca. 130k points per frame at 15 fps}                                     \\ \hline
Conditions                               & Target Bitrates                       & \multicolumn{1}{l|}{7 Mbps}         & \multicolumn{1}{l|}{14 Mbps}      &                \\ \hline
                                         & Streaming Conditions                  & \multicolumn{1}{l|}{Network Adaptive} & \multicolumn{1}{l|}{Tiled Adaptive} & Uncompressed   \\ \hline
\end{tabular}
\end{table*}

\subsection{Point Cloud Delivery}
\subsubsection{Compression Standards}
Point cloud compression has received significant research attention in recent years with the launch of two new MPEG compression standards ~\cite{Pablo:MPEGPCC}. The V-PCC standard codec for dynamic point clouds that projects point clouds geometry and attributes onto separate 2D patches that are them packed into video tracks along with the occupancy. These video tracks are then encoded using legacy video codecs making this approach suitable for relatively dense and uniform distribution of points. The G-PCC standard codec uses an octree space partitioning structure to code geometry and can be optionally combined with an additional surface reconstruction step using the TriSoup approach ~\cite{pavez:soup}. G-PCC also includes several modules for attribute coding, the lowest complexity coder uses the Region Adaptive Hierarchical Transform (RAHT) ~\cite{Queiroz:RAHTOriginal}. This codec is targeted at irregular sparse distributions of points making it suitable for live captured point clouds. However, both codecs introduce high complexity encode making them unsuitable for real-time communication. At the start of the MPEG standardization activity an anchor codec proposed by Mekuria et al. ~\cite{mekuria:tele} was introduced. This codec utilizes octree occupancy to code geometry and scans attributes to map them to a 2D grid to maximize correlations amongst co-located points and encodes them with legacy JPEG image compression. This approach offers low encode and low decode complexity making it suitable for real-time framerate sensitive applications such as VR remote communication. \textbf{In this work}, we use the anchor codec to encode point cloud tiles at multiple quality levels in real-time before streaming.

\subsubsection{Adaptive Streaming}
Initial works on adaptive streaming of point clouds utilized entire point cloud objects as the basic unit of bandwidth allocation in scenes containing multiple point cloud objects. Hosseini et al. ~\cite{christian:mmsys18:pcstreaming, Hosseini:PCC-DASHExtended} propose DASH-PC for dynamic adaptive view aware point cloud streaming. They propose three spatial subsampling techniques to create multiple representations of point cloud objects in a scene. The density of each object representation is used by the client for bitrate allocation based on human visual acuity. Hooft et al. ~\cite{Jeroen:MM:PCAdaptiveStreaming} propose PCC-DASH, a standards compliant means for HTTP adaptive streaming. They present three heuristics based on the users viewport and distance to the object to allocate bitrate to different objects in the scene. Different ranking metrics and bitrate allocation heuristics had to be selected for different scenes and user navigation paths. 

Another approach used in previous work is to split each point cloud object into tiles that are then used as the unit of bandwidth allocation. Park et al. ~\cite{phil:mmsys18:pcstreaming} define a utility per tile based on the user's proximity, point cloud surface quality and display device resolution. To account for interactions, they propose a window-based design for the Client Buffer Manager with greedy utility maximization. This type of rasterization and pixel occupancy based approach is not suitable as computing this at every frame in computationally expensive. He at al.~\cite{He:Hybrid} propose view-dependent streaming over hybrid networks. Each point cloud frame is projected onto the six faces of a bounding cube, with a color and a depth video created per face. The videos are transmitted using digital broadcasting. The user can request videos that correspond to particular faces of the cube in high quality from the edge node of a bidirectional broadband network, reconstructing the point cloud from the downloaded depth and color videos at the receiver end. This approach requires a redundant extra reconstruction step at each receiver. We instead perform the reconstruction on the sender side in order to generate a self view to embody the user and transmit the reconstruction to all receivers. Li et al.~\cite{Li:resourceallocation} propose a joint communication and computational resource allocation framework to stream and decode pre-recorded point clouds. They also propose a QoE metric to guide tile selection based on the users viewport, distance to tile and available quality levels. Lee et al.~\cite{kyungjin:groot} propose GROOT a real-time streaming system to reduce decoding overhead by dividing the point cloud into cells defined by the leaf nodes of an octree represented in a parallel decodable tree. Han et al~\cite{Bo:Vivo} propose ViVo using a similar approach and employ machine learning models to predict viewport movement. Liu et al.~\cite{zhi:fuzzystreaming} follow a similar approach, they include an uncompressed base layer and use fuzzy logic based quality selection. This type of approach using use the leaf nodes of the octree as an enhancement layer is currently not suitable for real-time systems as it adds an extra surface orientation estimation step that introduces additional delays in the pipeline. Subramanyam et al. ~\cite{cwi:mm20} build on the ideas presented in PCC-DASH ~\cite{Jeroen:MM:PCAdaptiveStreaming} to tile point cloud content using low complexity surface estimation suitable for frame rate sensitive real-time applications. They performed an objective quality evaluation using prerecorded navigation paths and image distortion metrics. \textbf{In this work}, we build on their approach and create tiles based on surfaces visible to multiple depth sensors and estimate their orientation using the transformation matrix associated with each sensor.

\begin{figure*}[h!]
    \centering
    \includegraphics[width=0.8\textwidth]{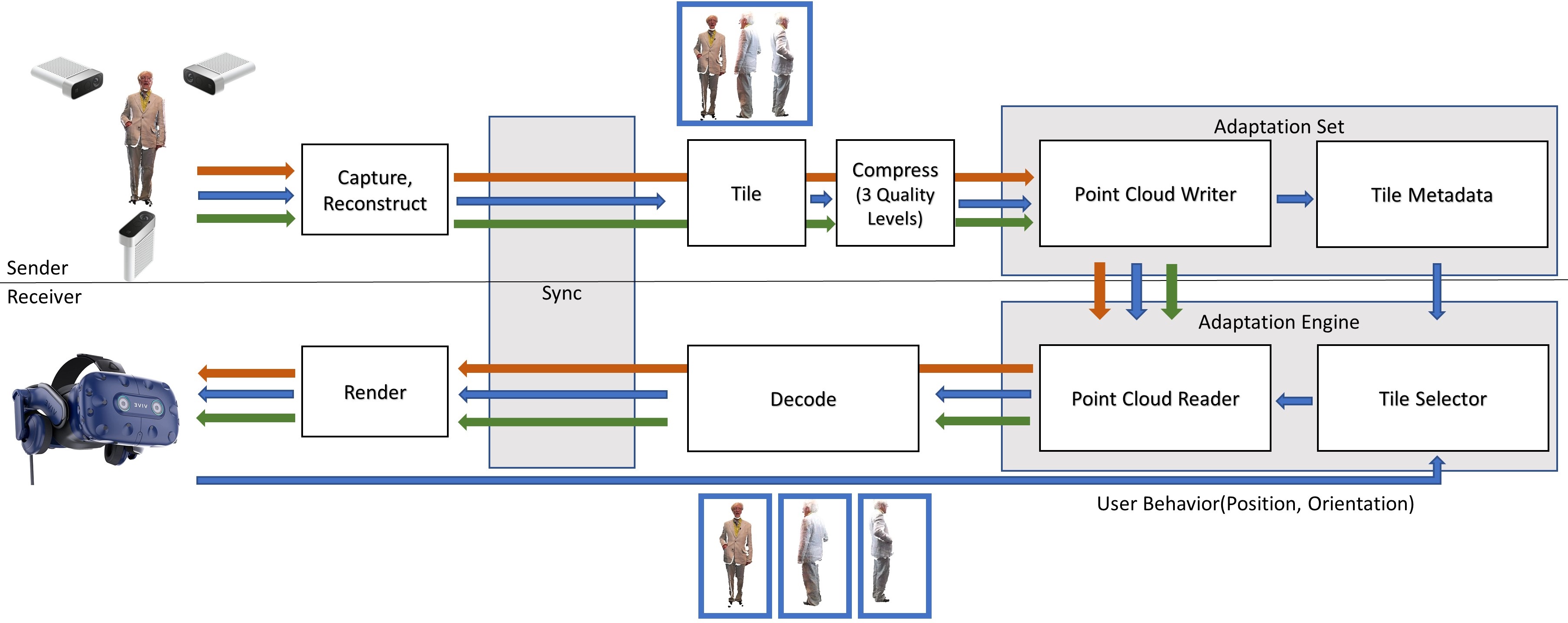}
    \caption{Architecture and Data Flow for Baseline (Orange), NA (Green) and TA (Blue) Streaming}
    \label{fig:arch}
\end{figure*}
\subsection{VR Communication Evaluation}
Quality assessment for remote communication is usually conducted using subjective user studies that are either passive or active. Passive tests involve asking users to rate prerecorded clips of content. this approach to evaluation is more suited for standardized testing of codecs with offline content and has limited ecological validity in remote communication~\cite{ITUTP1301,Marwin:videoconferencingQoE}. 
Active tests involve multiple remote participants being placed in an interactive live communication system. The International Telecommunication Union published recommendations to define evaluation methods for quantifying the impact of terminal and communication link performance on point-to-point audiovisual communication~\cite{ITUTP920}. The recommendation contains sample tasks such as name guessing, story comparison, picture comparison, object description and building blocks. Schmitt et al.~\cite{Marwin:videoconferencingQoE} utilize the building blocks task to develop and evaluate personalized quality of experiment metrics for multiparty video conferencing at varying bitrates. Smith et al.~\cite{Oculus:EmbodiedVRcommunication} compare face-to-face communication with embodied and unembodied remote VR communication. They propose a task involving negotiating apartment layout and furniture placement based on blueprints. Li et al.~\cite{jie:photosharingCHI} compare face-to-face, videoconferencing and Facebook spaces VR communication in the context of  a photosharing task. They found that Facebook spaces is able to closely approximate face-to-face photosharing. In general, these methods have been used to compare VR remote communication with other technologies and with face to face communication. The tasks proposed either focus on the audio quality or rely on external objects for the evaluation. \textbf{In this work}, we focus on evaluating adaptive streaming within VR remote communication. We define a new visually focused training task where participants are taught a neck exercise from a trainer and are asked to perform the exercise in order to complete the task. 

In order to evaluate communication, several questionnaires have been proposed in the literature. Toet et al.~\cite{TNO:HMSCQ} propose the holistic mediated social communication (H-MSC) framework and associated questionnaire to evaluate the experience of spatial presence as well as social presence. The framework is general enough to be used for any mediated social communication system. Slater et al~\cite{Slater:DepthOfPresence} and Witmer et al.~\cite{Witmer:PresenceQuestionnaire} have proposed two popular questionnaires aimed at measuring presence in virtual environments. Kangas et al.~\cite{Kangas:TaskRelated} present a pragmatic task related questionnaire that they use to evaluate VR interaction techniques in a rigid object manipulation task. Li et al.~\cite{jie:photosharingCHI} propose a social VR questionnaire that evaluates Quality of Interaction/Communication \textit{(QoI)}, Presence/Immersion and Social Meaning. \textbf{In this work}, we combine the QoI part of this questionnaire along with visual quality questions from~\cite{ITUTP920}, and task related questions from~\cite{Kangas:TaskRelated}.

\begin{figure*}
    \centering
        \begin{subfigure}
    \centering
    \includegraphics[width=0.08\textwidth]{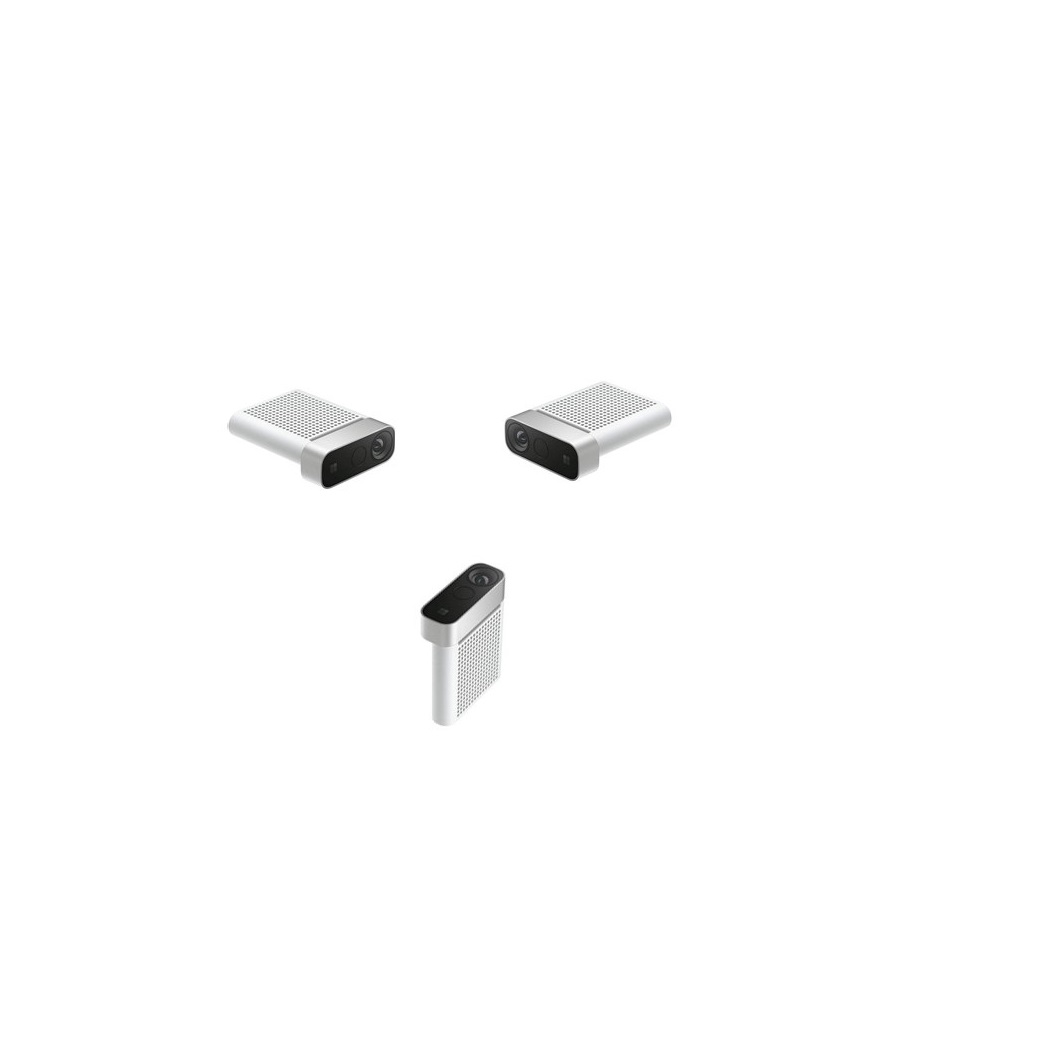}
    \end{subfigure}
    \begin{subfigure}
    \centering
    \includegraphics[width=0.2\textwidth]{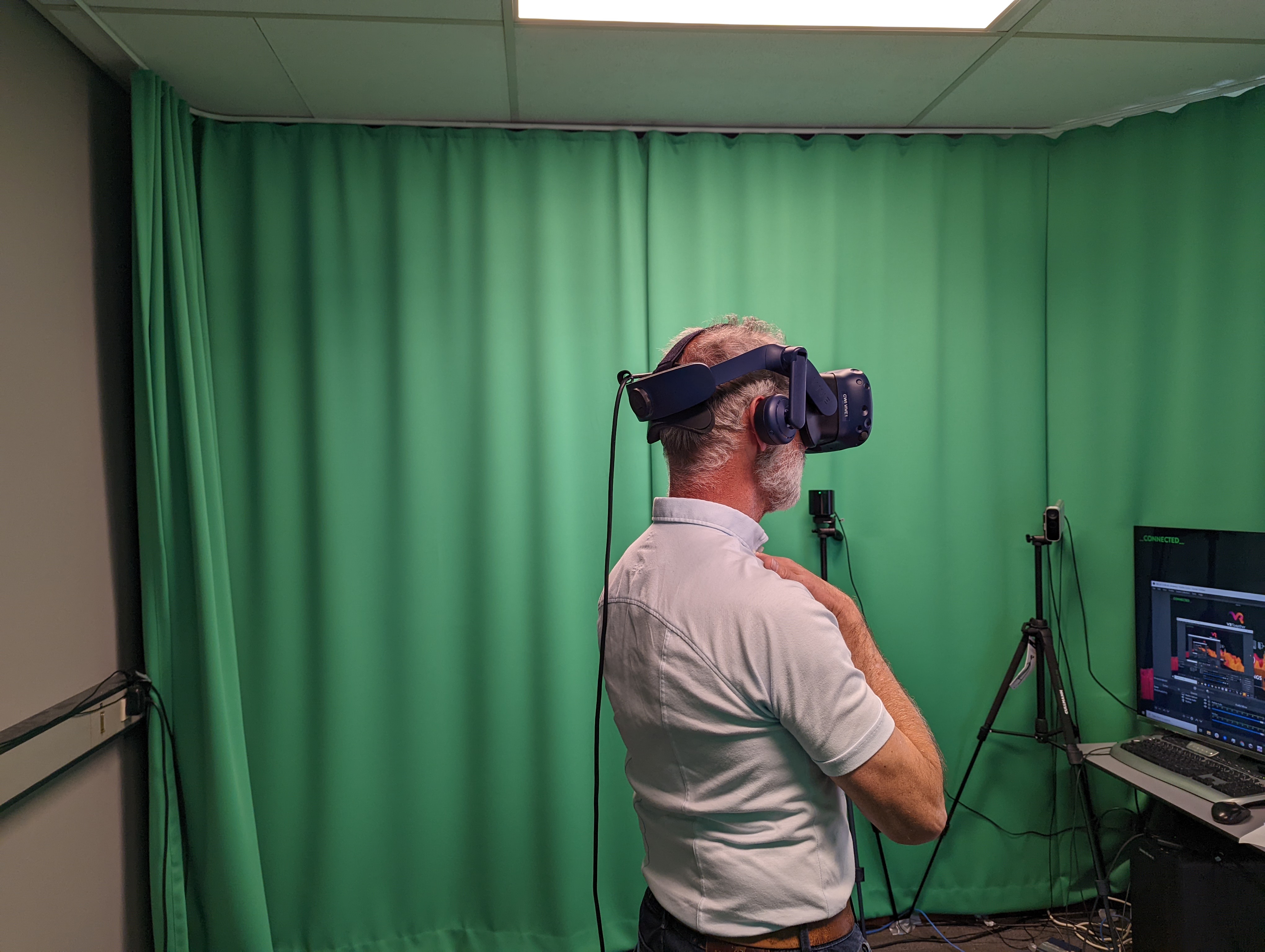}
    \end{subfigure}
    \begin{subfigure}
    \centering
    \includegraphics[width=0.2\textwidth]{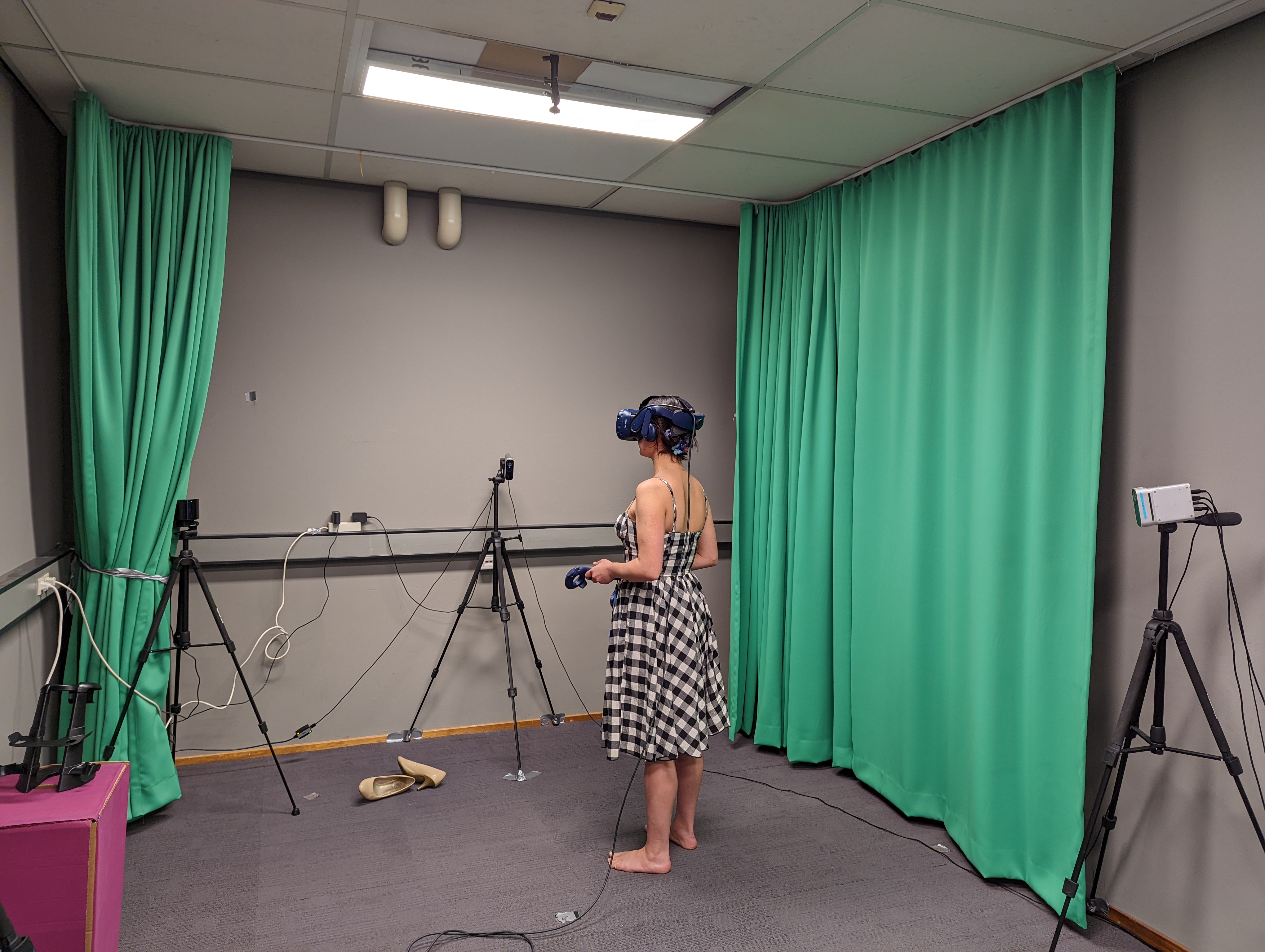}
    \end{subfigure}
        \begin{subfigure}
    \centering
    \includegraphics[width=0.2\textwidth]{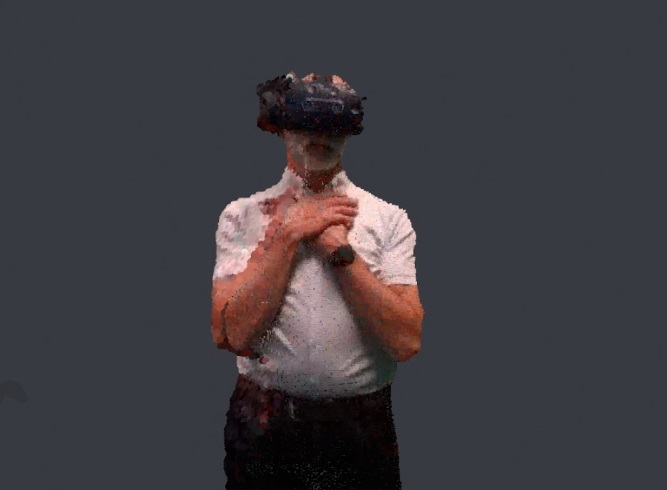}
    \end{subfigure}
    \begin{subfigure}
    \centering
    \includegraphics[width=0.185\textwidth]{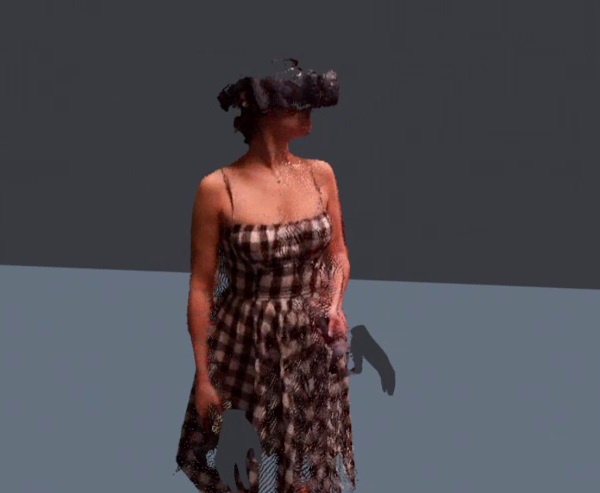}
    \end{subfigure}
    \begin{subfigure}
    \centering
    \includegraphics[width=0.08\textwidth]{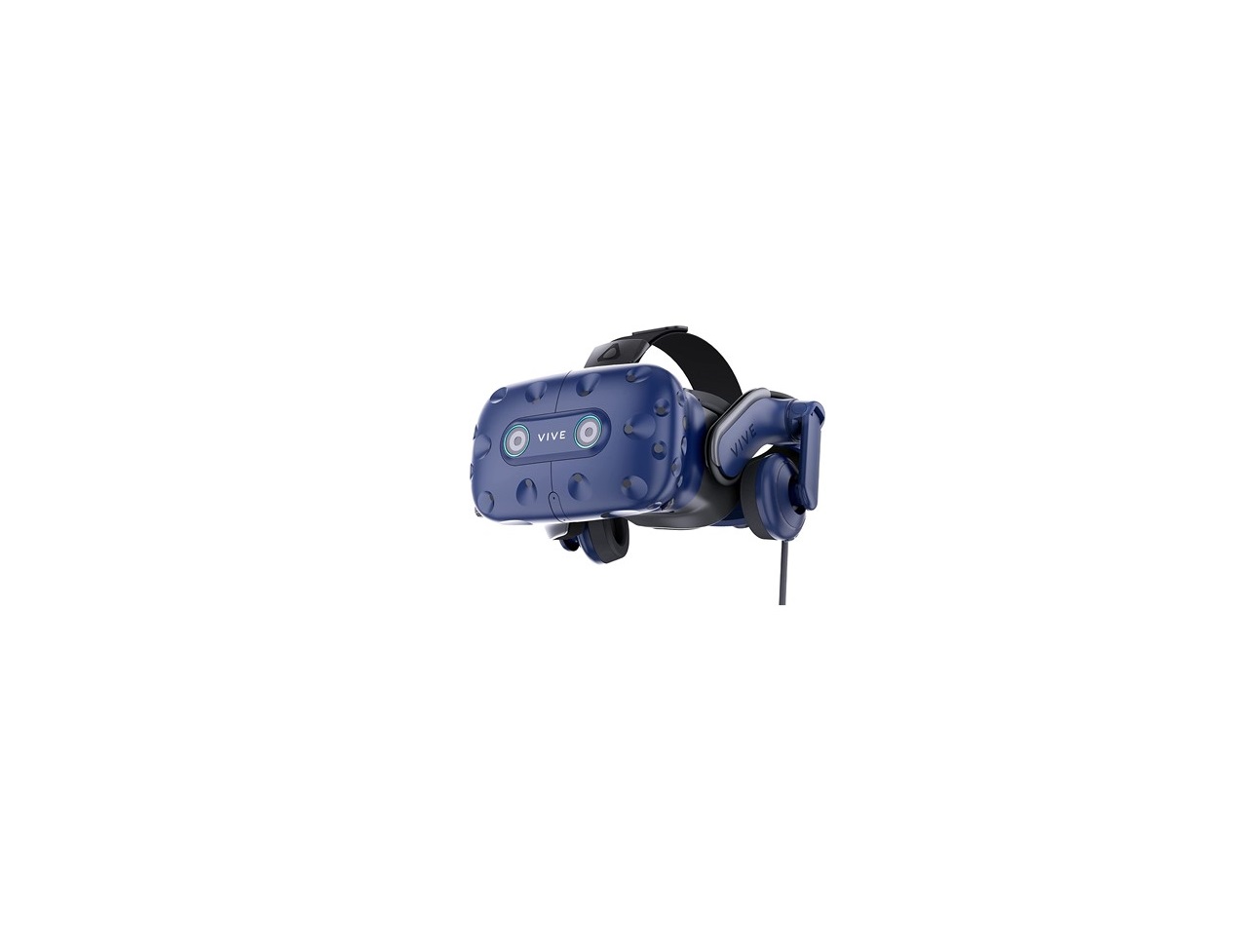}
    \end{subfigure}
    \caption{Capture Nodes and Real-Time Point Cloud Reconstructions}
    \label{fig:participants}
    \vspace{-0.2cm}
\end{figure*}

\section{Study Design}
\label{sec:studyDesign}
\subsection{Experiment Task}
The goal was to define a task that could be used to evaluate QoI, visual quality and task experience in VR remote communication. We needed to facilitate a conversation with a fixed general outline and with a focus on the visual aspect of communication. We considered the tasks presented in ITU-T P.920~\cite{ITUTP920}, the building blocks task presented in~\cite{Marwin:videoconferencingQoE} as well as the photosharing task presented in~\cite{jie:photosharingCHI}. We found that these tasks are either heavily dependent on the audio quality (story comparison, name-guessing and object description task) or required external objects that had to either be live captured along with the user or digitally represented in the virtual world with appropriate interaction tools (picture comparison, building blocks, photo-sharing). These objects could occlude and distract from the visual quality of the user reconstruction. In order to assess the impact of tiled adaptive point cloud streaming, we chose to keep the audio quality consistent across sessions and manipulated the quality of the point cloud representation of the participants. Based on pre-trials with seven colleagues, we selected a training task where participants were asked to first learn and then perform a neck exercise. This allowed for a fixed general outline of the conversation and focused on the visual representation of the remote participant. The neck exercises were found not to induce severe motion sickness as validated in section~\ref{sec:subjectiveAnalysis} and provided coherent results. 
In our experiment design, we use a confederate user as the trainer. In this approach, all participants (trainees) in the same condition are always paired with the same trainer. This allows us to focus on the individual as the unit of study and we mitigate social context as a confounding variable. In addition, we isolate the basic elements of communication by attempting to hold constant the behavior of one conversation partner. In order to adhere to the recommendations on using confederate users~\cite{Kuhlen:conferederateRecommendations}, we used an asymmetric communication channel. The trainers are always shown the same quality point clouds of the trainee (octree depth 9) in order to ensure that the confederate users are as naive as possible to the current experimental condition. In addition the trainers were not briefed on the hypothesis associated with each experimental condition. Confederate behavior could be scripted as these users were always the initiators and addressers as they provided instructions on how to perform the neck exercise. We then evaluate the quality ratings only from the point of view of the trainees in line with the recommendations in ITU-T P.1301~\cite{ITUTP1301}.
\subsection{Adaptive VR Remote Communication System}
\subsubsection{System}
The overall architecture and dataflow of the real-time VR remote communication system is shown in figure \ref{fig:arch} with baseline, TA and NA streaming. This is an extension of the system presented in~\cite{jack}. Apart from the point cloud delivery pipeline described here the actual implementation also contains an audio delivery pipeline and a module for session management. The point cloud capture, and codec modules are implemented in C++, the other modules in C\# with overall control in the Unity game engine.

The capture module interfaces with three Azure Kinect Depth sensors as shown in figure~\ref{fig:arch}. The sensors are calibrated in advance to generate the transformation matrix with intrinsics to combine color and depth as well as extrinsics to bring all sensors to a common coordinate system. The color and depth images from the sensors are then transformed and fused in order to reconstruct the point cloud. We set a target capture framerate of 15 fps based on pre-tests with colleagues as this was the maximum achievable framerate at an acceptable baseline quality. The point cloud generated is first sent directly to the renderer in order to generate a self view to embody the participant. 

In baseline uncompressed, the point cloud is serialized and sent directly to the writer to forward to the receiver. For NA, the point cloud is sent to the encoder where it is compressed to three quality levels and sent to the writer to prepare the adaptation set with the associated encoded size.

For TA, the point cloud is split into tiles based on the contributing sensor. Each tile contains an orientation vector that is derived from the transformation matrix of the sensor and the centroid of it's bounding box. The tiles are then fed to the compression module that launches encoders in parallel for each tile and quality level. In this way we create an adaptation set with multiple representations for each tile. In addition we prepare a tile meta data structure that contains information on the number of tiles, meta data for each tile, available quality levels and the associated encoded size. 

At the receiver, for baseline uncompressed the point cloud is sent directly to the renderer. For NA, the adaptation engine selects the highest possible quality within the available bandwidth budget. This is then decoded and sent to the renderer. We apply the bitrate budget per frame based on the target capture framerate of 15 frames per second (fps).

For TA, the adaptation engine utilizes both the tile metadata from the sender and the receiver's interactions in the system in terms of viewport position and orientation to select an appropriate representation for each tile within the available bandwidth budget. The tiles are then decoded in parallel and sent to the synchronizer. The synchronizer module was developed to playback tiled point cloud sequences with tiles of varying sizes and quality in a consistent manner. The primary goal of the synchronizer is to playback tiles of the same frame together with a secondary goal of playing back frames at the right time to match the received frame rate. The point clouds are then sent to the renderer.

Finally, the renderer stores the point locations and colors on a vertex buffer and draws procedural geometry on the GPU. Points are rendered as camera facing quads with a fixed offset based on the selected quality level. 

\subsubsection{Tiling and Tile Selection}
We use a modified version of the approach proposed in~\cite{cwi:mm20} in order to create tiles. The point cloud is split into tiles based on the contributing sensor. We use the forward vector of the contributing sensor to estimate the orientation of the tile surface. We also compute the centroid of the bounding box of the tile. Each tile $T_{i}$ has an orientation $\vec T_{i}$ and a bounding box centroid $T_i^{(bc)}$. The current viewport $V$ of the user is defined by a position $V^{(pos)}$ and an orientation $\vec V$. The utility of each tile is calculated based on the following formula: 

\begin{equation}
u(V, T_{i}) = 
\begin{cases}
 \big|\vec T_{i} \cdot \vec V \big|,  & \text{if } d(V^{(c)},  T_i^{(bc)}) < d_{max} (V^{(bc)}, T_j^{(c)})\\
- \big|\vec T_{i} \cdot \vec V \big|, & \text{otherwise.}
\end{cases}
\end{equation}

We use the absolute value of the the dot product to identify surfaces directly facing the user. In addition, to account for the position of the user we ensure that the two tiles closest to the user always have a positive utility.

In the next step the calculated utility is used to allocate the bandwidth budget to each tile. In this work, we use the allocation strategies presented in~\cite{subramanyam2020comparing,Jeroen:MM:PCAdaptiveStreaming}. Based on pre-tests with colleagues we found that uniform bit rate allocation achieved a higher median score and a lower spread of scores. We use the utility to rank the available tiles. The quality of each tile is then increased one step at a time in order of this ranking. 

\subsection{Experiment Conditions}
The primary components determining the point cloud quality are the capture sensors, the codec configurations or adaptation set, target bitrate, streaming condition  and the display device used. In order to evaluate adaptive streaming we vary the streaming condition and target bitrate. The other factors do not change dynamically over a session and are uninteresting for streaming optimization. They are held constant across all participants. In addition the audio quality is also maintained at the same level across all experimental sessions using the Ogg container format and the Speex codec with an ultra wide band sampling rate. In order to conduct a pre-test and set the experimental conditions we used the dataset published by Reimat et al~\cite{nacho:cwipcsxr} sub-sampled to three cameras (1,5,6) as this most closely resembles our capture setup. Based on pre-tests on encode time and captured point count we use three codec configurations on the MPEG anchor codec shown in table~\ref{tab:systemSetup}. All codec configurations were encoded at JPEG QP 75. We also set two target bitrates exclusively for remote user point cloud reconstructions at 7Mbps and 14Mbps based on the approximate bandwidth requirement for full point clouds using the two highest quality levels from the pre-tests. We selected three neck exercises that were to taught to all participants. In the experiment, we evaluate three streaming conditions: baseline uncompressed, NA and TA.

\subsection{Experiment Design and Protocol}
With three neck exercises and three streaming conditions we used a Greco-Latin square design to randomize and counter balance the different levels of each variable. In order to avoid fatigue we separated the target bitrates so each participant only took part in three sessions at a fixed target bitrate. We recruited two confederate users to play the role of trainer for each target bitrate. We recruited 16 and 17 participants for the two target bitrates of 7Mbps and 14Mbps. 

Upon arrival participants were led to the experiment room and were briefed about the purpose of the study, after which they were asked for written consent for data gathering. Participants were asked to provide some background information on themselves and to take a Ishihara test~\cite{ishihara} for color perception. Participants were then asked to fill the simulator sickness questionnaire before starting and again after each experiment session. We then conducted a brief training session where participants were shown the highest and lowest available quality of the remote user point cloud to serve as an anchor. Participants were then taught how to use the HMD controllers to teleport and navigate the virtual space. Participants were informed that during the experiment they are free to move about the virtual space. Participants then entered the first session, in each session there was a brief introduction by the trainer, a training stage where the trainer demonstrated the exercise technique and finally a performance where participants were asked to repeat the exercise three times in order to complete the session. Each session took 2 to 3 minutes to complete. Participants were asked to fill in a questionnaire to report their experience after each session. Participants completed the experiment in ca. 30 minutes.

\subsection{Experiment Setup}
Participants took part in the study in a separate room from the  trainer. Each room had a workstation, an HTC Vive Pro Eye HMD (with controllers and base stations) and three azure kinect depth sensors. The configuration of the setup used by participants is shown in table~\ref{tab:systemSetup}. Figure~\ref{fig:participants} shows the setup used to capture each user and the resulting point cloud. The two labs were connected using a dedicated gigabit ethernet connection in order to control the connection quality for the duration of the study. Each of the azure kinect cameras were set to use a depth mode of NFOV unbinned with a resolution of 640x576 and the lowest supported color resolution of 1280x720. This was done as the color image is later mapped to the depth image in order to reconstruct the point cloud similar to the method proposed in~\cite{nacho:cwipcsxr}.

During each experiment session the system resource consumption was measured using the Resources Consumption Metrics (RCM) tool~\cite{Mario:RCMTool}. The conversation audio was recorded with audio only capture using the Open Broadcaster Software tool. The VR remote communication application recorded log files that contained performance information on each component including latency and framerate. 

\subsection{Research Questions and Data Collected}
After each condition participants filled in a questionnaire about the experience they just had. The first six questions were related to QoI from~\cite{jie:photosharingCHI} with a 5-point Likert scale to address \texttt{R1}. The next four questions were about the visual quality of the point cloud representation taken from~\cite{ITUTP920} with a 5-point ACR scale to address \texttt{R3}. The last five questions were about task related experience from~\cite{Kangas:TaskRelated} with a 7-point Likert scale to address \texttt{R2}. In addition, on the trainee node we record system resource consumption and log playback performance to address \texttt{R4}.

\section{Results}
\label{sec:subjectiveAnalysis}

\subsection{Performance Results}


\begin{figure}
    \centering
    \begin{subfigure}
    \centering
    \includegraphics[width=0.23\textwidth]{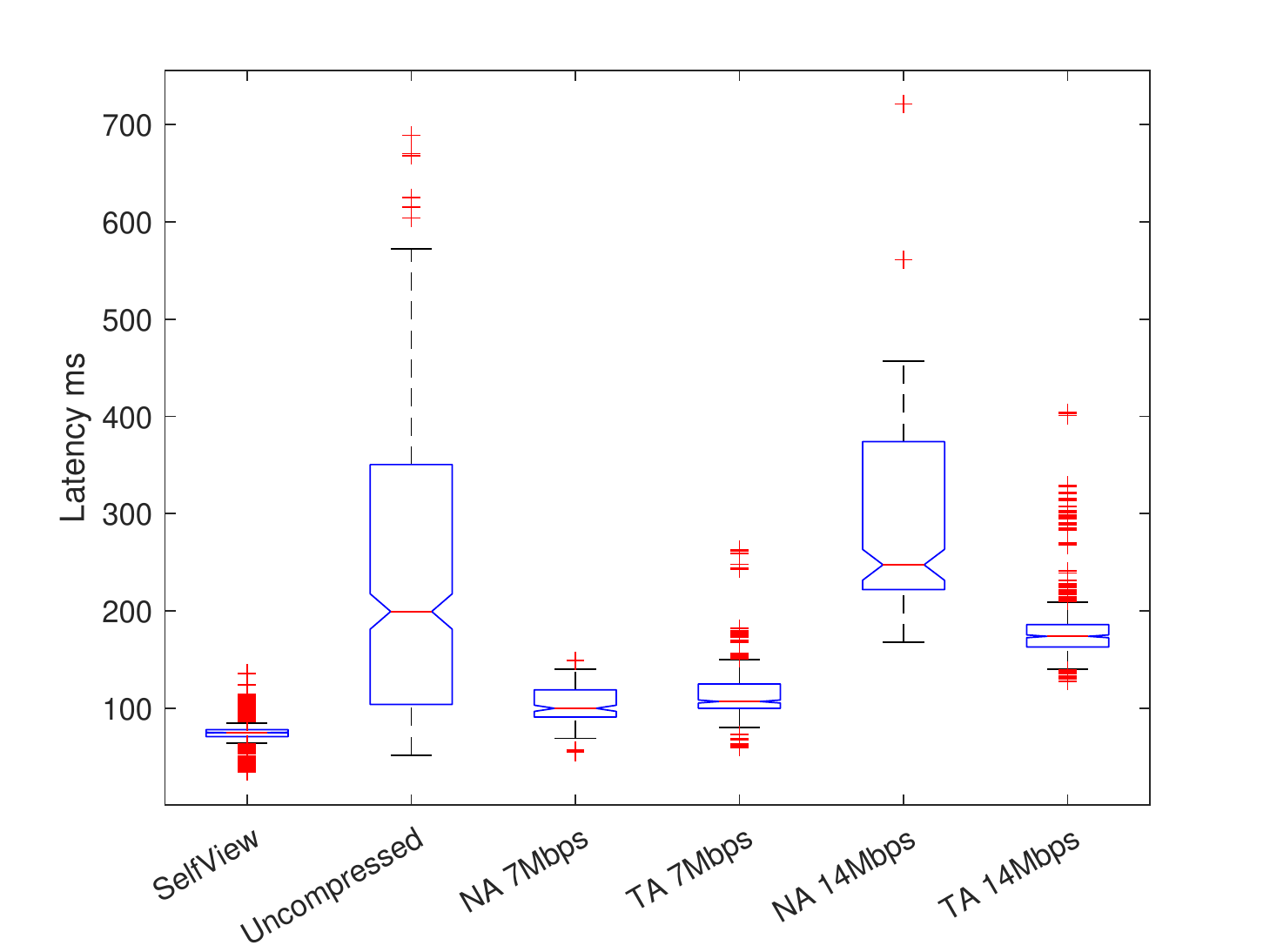}
    \end{subfigure}
    \begin{subfigure}
    \centering
    \includegraphics[width=0.23\textwidth]{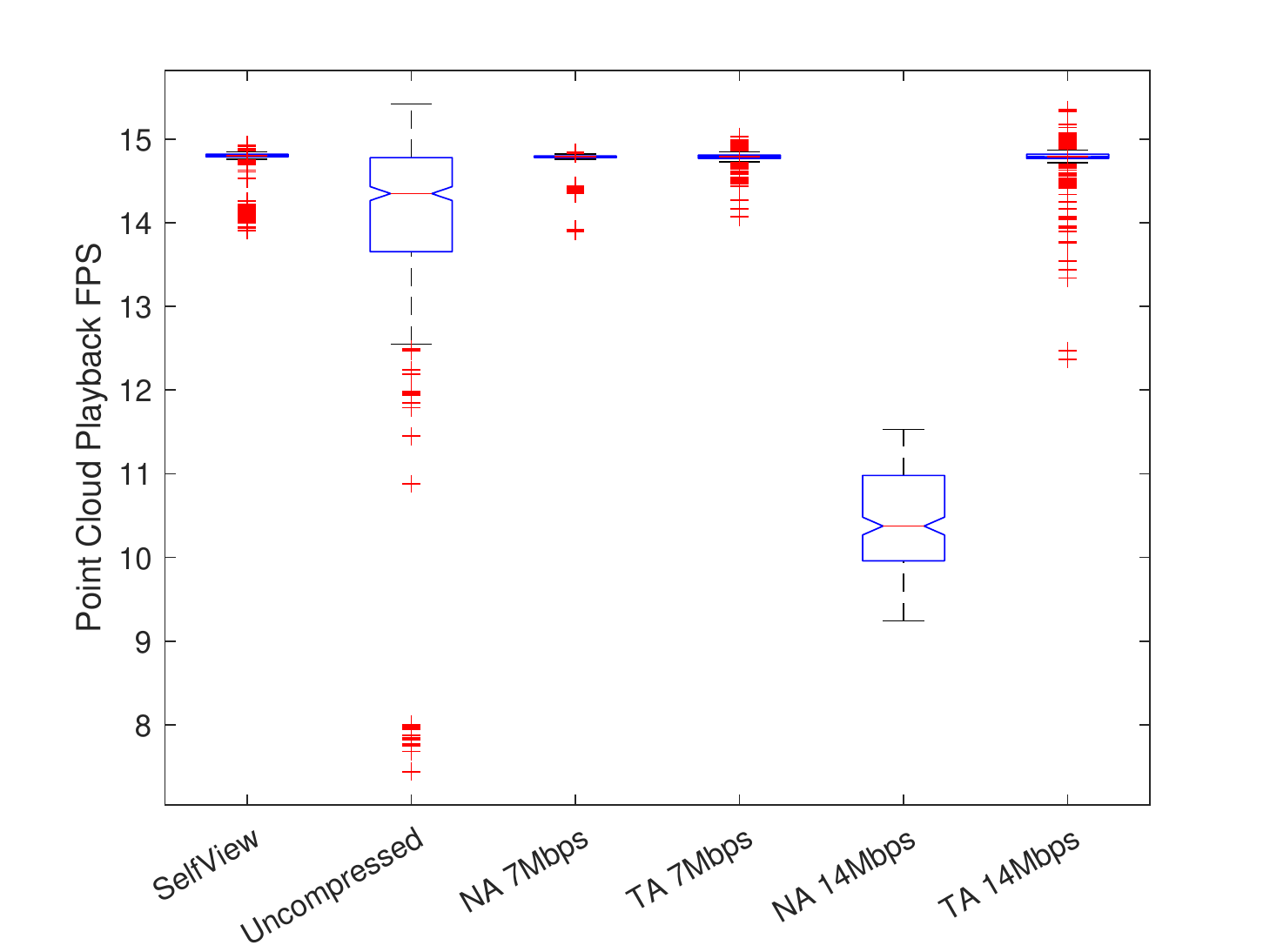}
    \end{subfigure}
        \vspace{-0.4cm}
    \caption{System Performance in terms of latency (ms) and framerate (fps)}
    \label{plot:latency}
        \vspace{-0.6cm}
\end{figure}


\begin{table*}[]
\caption{System Resource Consumption (1000+ samples each, mean values)}
\label{tab:rcm}
\begin{tabular}{l|l|l|l|l|l|}
\cline{2-6}
                                                  & Streaming Condition & Target Bitrate & CPU (\%)         & GPU(\%)         & Memory (MB)    \\ \hline
\multicolumn{1}{|l|}{Uncompressed}                & Baseline            & -              & 31.1\% (SD 7.8)  & 46.2\% (SD 6.8) & 3146 (SD 1423) \\ \hline
\multicolumn{1}{|l|}{\multirow{4}{*}{Compressed}} & Network Adaptive    & 7Mbps        & 49.7\% (SD 5.9)  & 45.2\% (SD 6.9) & 1025 (SD 112)  \\ \cline{2-6} 
\multicolumn{1}{|l|}{}                            & Tiled Adaptive      & 7Mbps        & 39.2\% (SD 10.7) & 46.9\% (SD 5.4) & 1088 (SD 94)   \\ \cline{2-6} 
\multicolumn{1}{|l|}{}                            & Network Adaptive    & 14Mbps       & 57.4\% (SD 4.6)  & 45.7\% (SD 4.8) & 1071 (SD 68)   \\ \cline{2-6} 
\multicolumn{1}{|l|}{}                            & Tiled Adaptive      & 14Mbps       & 39.2\% (SD 10.7) & 46.5\% (SD 4.7) & 1096 (SD 89)   \\ \hline
\end{tabular}
\end{table*}

We evaluate system performance based on resource consumption, framerate and latency. The machine used for the evaluation is described in table~\ref{tab:systemSetup}. Resource consumption was logged at the trainee node for each experiment session using the RCM measurement tool~\cite{Mario:RCMTool}. This tool is a windows application that allows for the capture of CPU, GPU and memory usage for a given process in a 1-second interval. The results are shown in table~\ref{tab:rcm}. As expected we see higher memory consumption for large uncompressed point clouds. At 7Mbps we observe ca. 10\% reduction in CPU usage on account of additional parallelization in encoding and decoding in TA as compared to NA with similar GPU and memory utilization. At 14Mbps we observe similar results with a 18\% reduction in CPU usage.

In addition, our VR application logs the capture to render latency and framerate while accounting for clock sync between the trainee and trainer node. The latency results are shown in figure~\ref{plot:latency}. The selfview latency describes the time needed for reconstructing and rendering only. The selfview is rendered with a median latency of 75ms across all experiment sessions. We observe the largest range of latency for baseline uncompressed streaming with the largest point clouds requiring ca. 300Mbps to transmit and render.  At 7Mbps we observe similar latency across the two streaming conditions. However, at 14Mbps we observe a 74ms increase in median latency for NA. This is caused by larger encode and decode times required for the highest quality point clouds in our adaptation set. In case of TA we have some performance gains due to parallel encoding and decoding of tiles and generally smaller point clouds decoded at the receiver.

The application runs at a near consistent 90 fps with motion reprojection. The point cloud target capture framerate is set to 15 fps based on the capability of the system. On the receiver end we observe a drop in median framerate to 10.4 fps for NA at 14Mbps caused by the encode and decode times required for the highest quality full point clouds in our adaptation set. For the remaining streaming conditions we generally observe similar performance of ca. 15 fps point cloud playback with a larger range for uncompressed point clouds as shown in figure~\ref{plot:latency}. To summarize, we observe significant gains in playback performance (framerate and latency) by employing TA with lower CPU usage as compared to NA to address \texttt{R4}.

\subsection{Subjective Results}
\subsubsection{Quality of Communication}
In this section of the questionnaire we included the QoI questions from~\cite{jie:photosharingCHI}. These questions are meant to assess four types of experience: (1) feeling understood (2) engaging in conversations (3) sensing other's emotion and (4) feeling comfortable in the environment. The overall QoI scores are obtained by adding up the scores for each of the six questions. We split the analysis for each target bitrate of 7Mbps and 14Mbps.

\begin{table}[h]
\centering
\caption{{Pairwise post-hoc test QoI and streaming condition at 7Mbps}} 
\label{table:qoi7}
\vspace{-0.1cm}
\begin{tabular}{@{}l|rrr@{}}
\toprule
Streaming Condition            & $Z$    & $p$               & $r$    \\ \midrule
NA -- TA            & -2.2340 & 0.0025 & 0.3950 \\
NA -- TA         & -2.6410 & 0.0083           & 0.4670 \\
TA -- Baseline        & -1.7230 & 0.0849           & 0.3050 \\ \bottomrule
\end{tabular}
\end{table}

For the 7Mbps case a Shapiro-Wilk normality test issued on the entirety of the scores indicates that they do not follow a normal distribution ($W=0.9163$, $p = 0.003$). We use non-parametric statistical tools to perform an exploratory data analysis and check if statistical differences could be found amongst the different streaming conditions. To compare the QoI across the different streaming conditions, we first conduct a Friedman's test to check if any groups exist with statistically significant differences ($\chi^2 = 12.04$, $p = 0.0024$). We then conduct a Wilcoxon signed rank test with Bonferroni correction. The results are shown in table ~\ref{table:qoi7}

We observe statistically significant differences in two of the comparisons. TA out performs NA with a medium effect size ($r=0.39505$). As expected, baseline uncompressed streaming out performs NA with a large effect size ($r=0.467$). On the other hand, we do not observe statistically significant differences between TA and baseline uncompressed indicating similar performance in terms of QoI.

\begin{table}[h]
\centering
\caption{{Pairwise post-hoc test QoI and streaming condition at 14Mbps}} 
\label{table:qoi14}
\vspace{-0.1cm}
\begin{tabular}{@{}l|rrr@{}}
\toprule
    Streaming Condition           & $Z$    & $p$               & $r$    \\ \midrule
NA -- TA            & -3.3720 & \textless{}.001 & 0.3780 \\
NA -- Baseline         & -3.3310 & \textless{}.001          & 0.5710 \\
TA -- Baseline        & 2.1650 & 0.0304           & 0.3710 \\ \bottomrule
\end{tabular}
\end{table}

For the 14Mbps case, a Shapiro-Wilk normality test indicates that the scores are not normally distributed ($W=0.9484$, $p = 0.002$). In order to check if any of the groups exhibit statistically significant differences we run Friedman's test ($\chi^2 = 24.96$, $p < 0.001$). We then conduct a Wilcoxon signed rank test with Bonferroni correction. The results are shown in table \ref{table:qoi14}.

This time we observe statistically significant differences in all comparisons. TA out performs NA with a medium effect size ($r=0.3780$). Baseline uncompressed streaming out performs NA with a large effect size ($r=0.5710$) and out performs TA with a medium effect size ($r=0.3710$). In general, we observe that TA leads to statistically significant gains in terms of QoI with respect to NA across both bitrates to address \texttt{R1}. 

In order to validate the three exercises we used, we checked if they led to different QoI scores and we found no statistically significant differences using the Friedman test ($\chi^2 = 3.16$, $p = 0.206$) at 7Mbps and ($\chi^2 = 2.28$, $p = 0.3198$) at 14Mbps.

\subsubsection{Visual Quality}
In order to assess the visual quality we include a question about the visual quality of the trainer's point cloud representation. Participants were asked to indicate the quality on a scale from 1 to 5 (\textit{1-Bad}, \textit{2-Poor}, \textit{3-Fair}, \textit{4-Good}, and \textit{5-Excellent}). We analyzed these scores separately for each target bitrate. 

\begin{table}[h]
\centering
\caption{{Pairwise post-hoc test visual quality and streaming condition at 7Mbps}} 
\label{table:vq7}
\vspace{-0.1cm}
\begin{tabular}{@{}l|rrr@{}}
\toprule
    Streaming Condition           & $Z$    & $p$               & $r$    \\ \midrule
NA -- TA            & -2.5840 & 0.0098 & 0.4570 \\
NA -- Baseline         & -2.5550 & 0.0106          & 0.4520 \\
TA -- Baseline        & -1.2650 & 0.2059           & 0.2240 \\ \bottomrule
\end{tabular}
\end{table}
For the 7Mbps case a Shapiro-Wilk normality test issued on the scores indicates that they do not follow a normal distribution ($W=0.8106$, $p <0.001$). To compare the remote user visual quality across the different streaming conditions, we first conduct a Friedman's test to check if any groups exist with statistically significant differences ($\chi^2 = 9.8$, $p = 0.0074$). We then conduct a Wilcoxon signed rank test with Bonferroni correction. The results are shown in table ~\ref{table:vq7}.
We observe statistically significant differences in two pairwise comparisons. TA out performs NA with a medium effect size ($r=0.4570$). Baseline uncompressed streaming out performs NA with a medium effect size ($r=0.4520$). We do not observe statistically significant differences between TA and baseline indicating similar performance in terms of visual quality.

\begin{table}[h]
\centering
\caption{{Pairwise post-hoc test visual quality and streaming condition at 14Mbps}} 
\label{table:vq14}
\vspace{-0.1cm}
\begin{tabular}{@{}l|rrr@{}}
\toprule
    Streaming Condition            & $Z$    & $p$               & $r$    \\ \midrule
NA -- TA            & -2.8140 & 0.0049 & 0.4830 \\
NA -- Baseline         & -3.1400 & 0.0017          & 0.5390 \\
TA -- Baseline        & -2.1210 & 0.0339           & 0.3640 \\ \bottomrule
\end{tabular}
\end{table}

\begin{figure}
    \centering
    \begin{subfigure}
    \centering
    \includegraphics[width=0.23\textwidth]{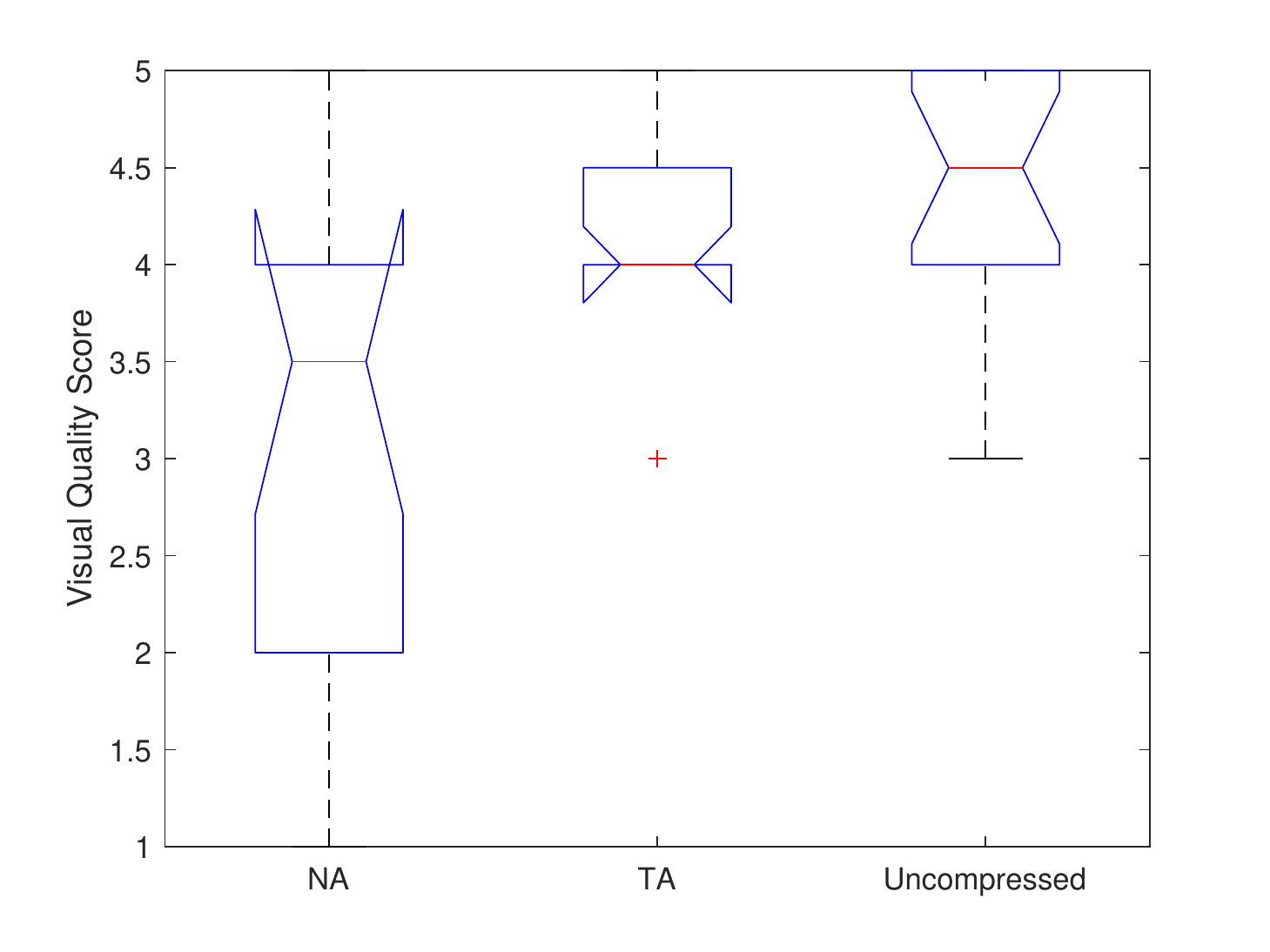}
    \end{subfigure}
    \begin{subfigure}
    \centering
    \includegraphics[width=0.23\textwidth]{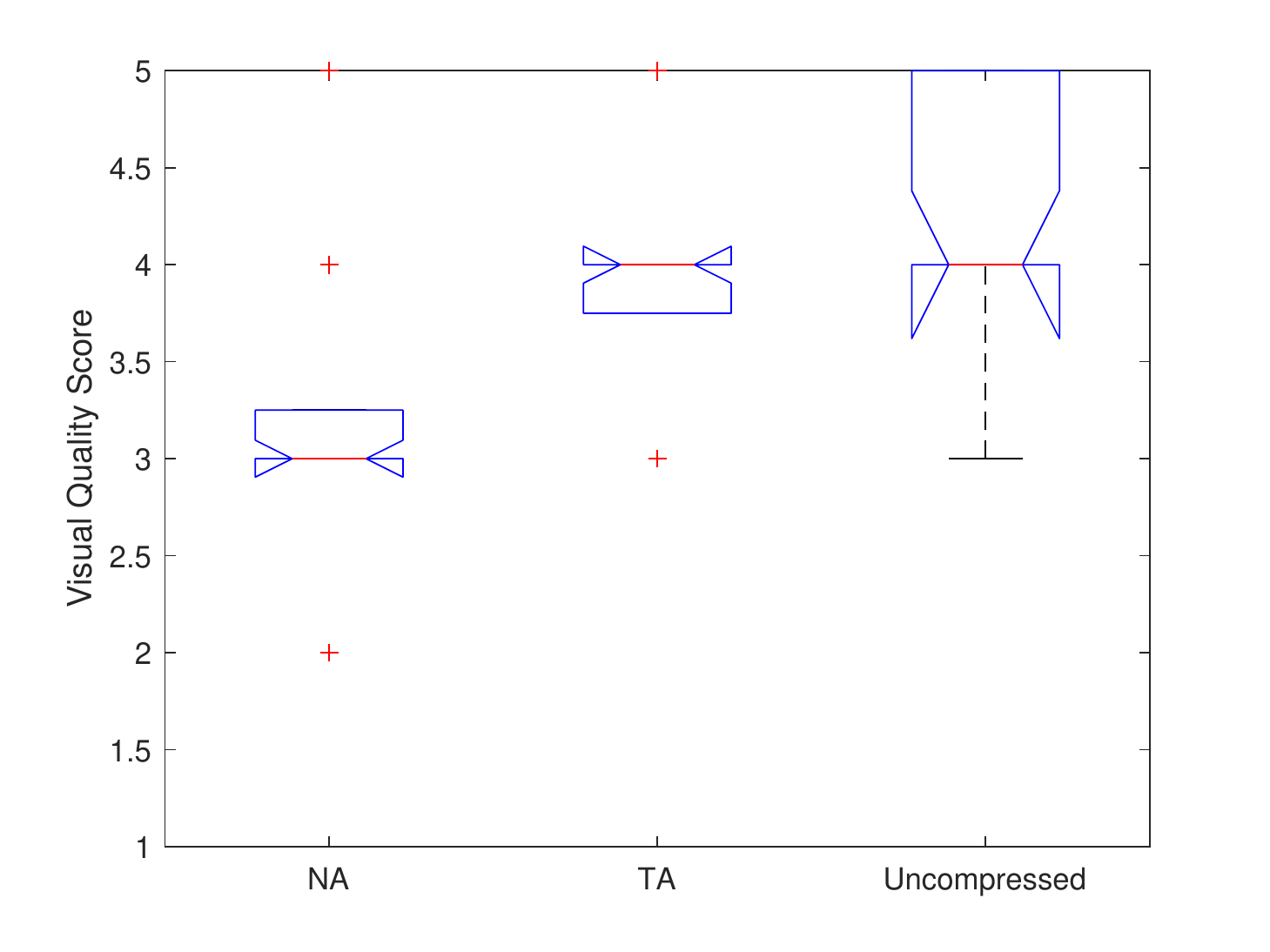}
    \end{subfigure}
        \vspace{-0.5cm}
    \caption{Point cloud quality at 7Mbps and 14Mbps}
    \label{plot:vq}
    \vspace{-0.4cm}
\end{figure}

For the 14Mbps case a Shapiro-Wilk normality test indicates the scores do not follow a normal distribution ($W=0.8622$, $p <0.001$). Friedman's test reveals that there are groups with statistically significant differences ($\chi^2 = 19$, $p < 0.001$). The results of the Wilcoxon signed rank test with bonferroni corrections are shown in table~\ref{table:vq14}.
This time we observe statistically significant differences in two of the pairwise comparisons. TA out performs NA with a medium effect size ($r=0.4830$). Baseline uncompressed out performs NA with a large effect size ($r=0.5390$). We do not observe statistically significant differences between TA and baseline uncompressed streaming indicating similar performance in terms of visual quality at 14Mbps. The distribution of the scores is shown in figure~\ref{plot:vq}. To address \texttt{R3}, we observe significant gains in perceived visual quality by employing TA over NA. At 14Mbps we even observe the same median score as baseline uncompressed streaming that required ca. 300Mbps.

\subsubsection{Task Related Experience}
For the task related experience questions we used the questionnaire presented in~\cite{Kangas:TaskRelated}. The questions are meant to assess the participants confidence in using the system and if the system was natural and easy to use while performing the task. We compute an overall score by adding up the five questions from this section. We split the analyses by target bit rate. A Shapiro Wilk normality test issued on the scores revealed that they are not normally distributed; ($W=0.9019$, $p = 0.0013$) for 7Mbps and ($W=0.9653$, $p = 0.014$) for 14Mbps. We then checked if there are any groups with statistically significant differences using Friedman's test ($\chi^2 = 2.1$, $p = 0.3504$) at 7Mbps and ($\chi^2 = 2.56$, $p = 0.278$) at 14Mbps. We found no statistically significant differences amongst the different streaming conditions at both target bitrates. This indicates participants were able to adapt their behavior to compensate for changes in the point cloud quality and were able to complete the task within the same time regardless. We observe no gains in task experience by employing TA (\texttt{R2}). This can be explained by the relative simplicity of completing our training task as compared to tasks used in other works~\cite{jie:photosharingCHI, Marwin:videoconferencingQoE, ITUTP920}.

\subsubsection{Simulator Sickness Questionnaire}
We calculate the total severity of cybersickness based on the SSQ questionnaire for all 33 participants across the three experiment sessions. We observe low post-exposure total severity scores for cybersickness, as defined in~\cite{ssqoriginal:kennedy, ssqvr:bimberg} ([mean, median, std] baseline: [10.54, 3.74, 18.22], after session 1: [9.63, 3.74, 14.88], after session 2: [9.41, 3.74, 16.14], after session 3:[11.22, 3.74, 16.95]). In general, no users reported severe symptoms after participating in any of the experiment sessions.


\section{Discussion}
\label{sec:discussion}

\subsection{Real World Experiments on Tiled Adaptive Streaming }
In our pre-trials we observed a higher median score and a lower spread of scores with uniform tile allocation. This is different from the results in~\cite{cwi:mm20} with a prerecorded dataset, where hybrid tile allocation was shown to yield a higher perceived quality. The point clouds used in that study were captured offline. They were dense and voxellized with ca. 1 million points per frame and the adaptation set comprised of 30 quality levels. In our study, we used real-time live captured point clouds with ca. 130K points with an adaptation set of 3 quality levels defined by octree depths. Further real world studies need to conducted to better understand the impact of these optimization techniques and acceptable quality differences amongst visible tiles for real-time reconstructions.

\subsection{Humanoid Point Cloud Considerations}
In general, tiled adaptive streaming techniques have received significant research attention for omnidirectional videos~\cite{fan:360vidSurvey,Zink:360Summary,360adaptation:petranjeli} and point clouds~\cite{petrangeli:AR3DObjects,Zhi:PCVidStreaming,jeroen:6DoFVRGeneral,Feng:SmartphoneVV}. However, further study is required for live real-time human point cloud reconstructions. Although the tiles are independently decodable, their quality cannot be optimized in isolation based on available bandwidth and viewport. Some participants reported that seeing artifacts in body extremities and in the face of the reconstruction was unpleasant. Further study into body part segmentation and quality perception is required to optimize tiling and tile selection strategies for humans engaging in conversation as compared to prerecorded content, objects and scenes.

 \subsection{Communication Evaluation Tasks}
 There is a need for new standardized tasks that can be used to evaluate VR remote communication. The existing ITU recommendations are insufficient to handle novel interaction techniques and immersive content inherent to VR communication. In this work, we utilized a neck exercise training task as it was more visually focused and the interaction was repeatable with a confederate trainer. Further study into other use cases and scenarios are required to evaluate emerging VR communication systems. To this end, ITU-T has recently launched a new activity~\cite{ITU-T-P.QXM} to develop assessment methods for extended reality meetings. 
 
 In our study, participants were trained on how to navigate the virtual space with a controller based teleport. During the session, we did not force participants to move in order to keep the interaction more natural. Future studies on VR remote communication should account for this trade-off. Movement within the scene is important to evaluate the visual quality of adaptation from different view angles but scripting or forcing these movements tends to break the flow of the interaction making it difficult to evaluate quality of communication.




\section{Conclusion}
\label{sec:conlcusionFutureWork}
In this paper, we presented a VR remote communication system with tiled adaptive real-time point cloud streaming using commodity hardware. We present an evaluation framework and a training task to evaluate the impact of adaptive streaming on QoI, visual quality and task related experience. Our system at 14Mbps was able to achieve similar visual quality as compared to uncompressed streaming at ca. 300Mbps. We also demonstrate statistically significant improvements to QoI as compared to traditional network adaptive streaming as well as improvements to playback performance with a 10\% to 18\% reduction in CPU usage.
\bibliographystyle{ACM-Reference-Format}
\bibliography{acmart.bib}










\end{document}